\documentclass[a4paper,11pt,twoside]{article}

\usepackage[utf8]{inputenc} 
\usepackage{color}
\usepackage{graphicx}
\usepackage{amsmath}
\usepackage{amssymb}
\usepackage[bottom=3cm,top=3cm,left=3cm,right=3cm]{geometry}
\usepackage{float}
\usepackage{url}
\usepackage[normalem]{ulem}

\newcommand{\be}{\begin{equation}}
\newcommand{\ee}{\end{equation}}
\newcommand{\ben}{\begin{equation*}}
\newcommand{\een}{\end{equation*}}

\usepackage{fancyvrb}
\usepackage{xfrac}

	\newcommand{\euler}[1]{\mathrm{e}^{#1}} 

	\newcommand{\uinc}{u^{\text{inc}}}

	\newcommand{\muller}{{M}}
	
	\newcommand{\vb}[1]{\ifmmode\mathbf{#1}\else$\mathbf{#1}$\fi} 

	\newcommand{\TS}{\text{TS}} 
	
\begin{document}

\markboth{Gonzalez, Lavia, Blanc, Maas, Madirolas}
{Boundary-element method to analyze acoustic scattering from a coupled swimbladder-fish body configuration}


\title{
Boundary-element method to analyze acoustic scattering from a coupled swimbladder-fish body configuration
}
\author{J. D. Gonzalez$^{1,2}$, E.F. Lavia$^{1,2}$, S. Blanc$^{1,2}$, M. Maas$^3$, A. Madirolas$^4$}
\maketitle

\begin{center}
\small
\smallskip
$^1$ \textit{{Acoustic Propagation Department. Argentinian Navy Research Office (DIIV), Laprida 555 Vicente Lopez, 
(1638) Buenos Aires, Argentina}}

$^2$ \text{UNIDEF (National Council of Scientific and Technical Research – Ministry of Defense)}

$^3$ \textit{Computational Simulation Center - CONICET, Godoy Cruz 2390 C1425FQD, Buenos Aires, Argentina}

$^4$ \textit{National Institute for Fisheries Research and Development (INIDEP). Paseo Victoria Ocampo 1. Escollera 
Norte. B7602HSA Mar del Plata. Argentina}
\end{center}

\begin{abstract}
A model for computing acoustic scattering by a swimbladdered fish with coupling to surrounding fish tissue that is
assumed to behave as a homogeneous fluid, is presented.
Mathematically, this corresponds to considering the problem of two penetrable scatterers immersed in a homogeneous medium,
one of which is wholly embedded in the other.
The model is formulated in the frame of boundary integral equations whose solution is achieved using the Boundary
Element Method (BEM) for a planar triangle mesh.
The numerical implementation is verified against benchmark solutions reported in the literature.
The model is then applied to a specimen of \textit{Merluccius hubbsi}, whose morphometry was determined by
CT scanning, for evaluating its forward and backscattering responses.
From the acoustic scattering viewpoint, the swimbladder is considered as a gas-filled object while the flesh constituting the 
fish body acts like a weak scatterer. 
The numerical results suggest the swimbladder and the fish body responses, when fully coupled, can lead to 
substantial differences with respect to the simplified models normally in use in the area of aquatic ecosystem research.
\end{abstract}

\clearpage

\section{Introduction}
\label{Intro}

Modelling of the acoustic scattering from a single fish is a problem which has been extensively studied 
over the past decades \cite{haslett1969target, gorska2003modelling, clay1994acoustic}. Since the swimbladder plays a fundamental
role from the acoustic viewpoint \cite{foote1980importance}, many models for the evaluation of scattering by individual fish, which 
exclusively consider  the swimbladder and neglect any other contribution, have been  reported 
\cite{francis2003depth,reeder2004broadbandExperimento}. The swimbladder has been modelled as a non-penetrable body with 
a pressure-release (Dirichlet) condition on its boundary or as a fluid (gas-filled) body with a canonical geometry 
(cylinder, sphere or spheroid) \cite{ding1997method,okumura2003acoustic}. This last case requires the solution of a 
substantially more involved transmission problem since the acoustic field inside the object representing the swimbladder 
is required. 

More realistic models explicitly include the contribution of the swimbladder and the surrounding body (fish flesh, 
bones and internal organs) by adding both contributions either coherently or incoherently. This kind of approach does 
not take into account the interaction between the aforementioned scattering components 
\cite{gorska2003modelling,clay1994acoustic,prario2015prolate} which, as shown in \cite{ding1997method}, is inadequate 
under certain conditions. 
In fact, the body of the fish and the swimbladder should not have to be considered separately 
but as a coupled acoustic system. 
On the other hand, in some other references \cite{gorska2003modelling,prario2015prolate,macaulay2013accuracy,perez2018numerical}, 
exact or approximate methods are applied to simplified geometries such cylinders and prolate spheroids which are used to model 
the scatterers. 

The Boundary Element Method (BEM) has also been widely used to predict scattering by fish 
\cite{francis2003depth,foote2002comparing}. Among its advantages, it can be pointed out the fact that this method can 
handle complex geometries and the solution space is one dimension lower than that the scatterer object 
since field values over surfaces instead of field values over volumes are required \cite{mallardo1998boundary}.
However, fish scattering predicted by BEM has mainly been applied to individual targets (i.e. 
considering either the fish body or the swimbladder) \cite{francis2003depth, okumura2003acoustic}.


Moreover, there are reported articles where both contributions, fish bones and swimbladder, are taken into account in 
acoustic scattering by fish, through the method of \textit{fundamental solutions} \cite{perez2018numerical}. The 
scattering from a fish with complex geometry, acquired through CT-scan technology, is computed in 
\cite{reeder2004broadbandExperimento}. Their authors consider two types of models. At first, only the swimbladder 
response is taken into account and the computations are carried out by the conformal mapping-based Fourier Matching 
Method (FMM) \cite{reeder2004acousticTeoria}, which is suitable exclusively for axi-symmetric bodies. Secondly, in order 
to take into account the swimbladder and fish body, they use the Kirchhoff Ray Model (KRM) 
\cite{clay1991low,clay1994acoustic}, where both contributions are coherently added. 

As far as the authors are aware, there seems to exist a relative void of studies that account for the interaction 
between the fish flesh and the swimbladder, without simplifying hypotheses about the shape of the bodies. This problem 
can be modelled by a BEM formulation that considers a double transmission scattering problem (i.e. a penetrable 
scatterer inside another). In this work a BEM approach is used for the double transmission scattering problem that is 
suitable to handle complex geometries represented by an ensemble of triangular facets but still keeping the 
simplification on the complex anatomy of a real fish by assuming homogeneous material properties within each scatterer 
volume. 

This paper is structured as follows. In Section \ref{modelformulation} the complete integral-equation 
formulation of the problem is provided. Section \ref{verifications} presents comparisons with benchmark solutions 
previously reported in the literature \cite{jech2015comparisons} as a means to get a model verification. In the Section 
\ref{application}  the model  is applied to a complex geometry acquired from computer tomography of the fish specimen 
\textit{Merluccius hubbsi}. Finally, the Section \ref{conclusions} summarizes the main conclusions of the work.

\section{BEM Model formulation}
\label{modelformulation}

\subsection{Acoustic problem and integral formulation}

\begin{figure}[tb]
	\centering
	\includegraphics[scale=0.5]{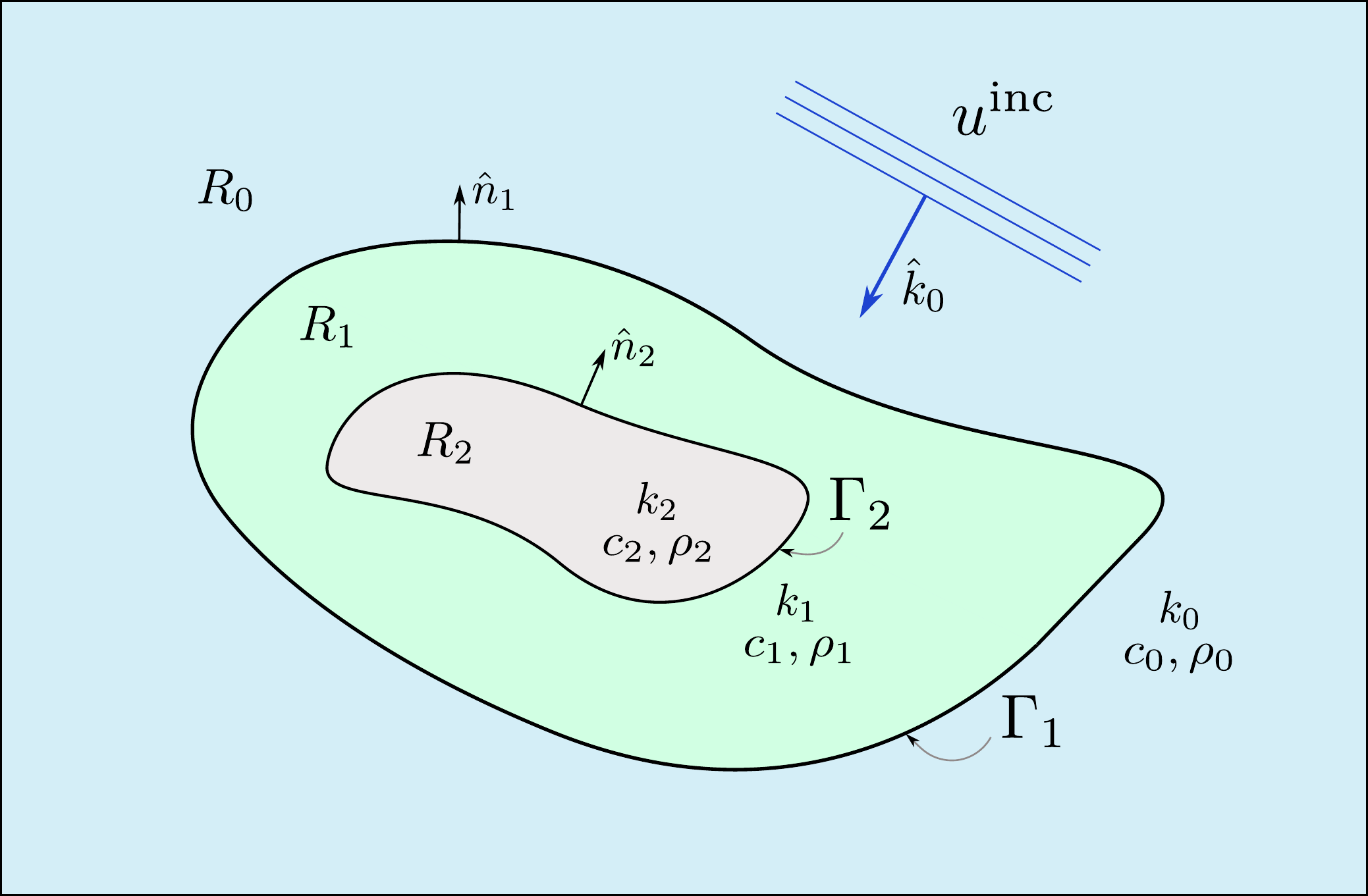}
	\caption{Scheme for the acoustic scattering of an incident field $ \uinc $ by two homogeneous 
	objects immersed in an unbounded domain, with boundaries $\Gamma_1$ and $\Gamma_2$ which define 
	three volumetric regions $R_i$ ($i=0,1,2$) with  physical properties $k_i, c_i, \rho_i$ 
	(wavenumber, sound speed and density, respectively).}
	\label{esquema}
\end{figure}

The problem of harmonic acoustic scattering by two fluid (i.e. acoustically penetrable) objects, 
one of which is wholly embedded in the other is schematically shown in Figure \ref{esquema}.
An incident acoustic pressure field $\uinc$, a plane wave propagating with frequency $\omega$ and direction $\hat{k}_0$, 
interacts with two penetrable objects delimited by boundary surfaces $\Gamma_1$ and $\Gamma_2$, whose respective exterior 
normals are $\hat{n}_1$ and $\hat{n}_2$.
These boundaries delimit three volumetric regions $R_i$ ($i=0,1,2$) whose physical properties $c_i, \rho_i$ 
(sound speed and density) determine the corresponding wavenumbers $k_i = \omega / c_i$. The region $R_0$ is the medium where 
the field $\uinc$ propagates and is the only one that is unbounded.

In each region $R_i$ the resulting pressure field $u_i$ is a solution of the scalar Helmholtz equation 
\begin{equation}
	(\nabla^2 + k_i^2) \: u_i = 0.
	\label{helmholtzpuro}
\end{equation} 
This field $ u_i $ is the complex valued space-dependent part of the sound pressure field in the time-harmonic case.
In the unbounded region $R_0$ the total field is $u_0 + \uinc$, where $u_0$ is the so called {\it scattered} field
which must satisfy a Sommerfeld radiation condition at infinity \cite{CK98}.

At the boundaries $\Gamma_j$ ($j=1,2$) the transmission boundary conditions must be fulfilled, 
i.e. the total field and its normal velocity, $v_n=-1/( i \omega \rho ) \: \partial_n u$
where $ i^2 = -1$, must be continuous across the interfaces $\Gamma_j$ between the regions.

For the acoustic problem under consideration, these continuity conditions lead to 
\begin{equation}
\begin{array}{rclr}
 	\left. \begin{aligned}
	\uinc(x) + u_0(x) &= u_1(x)  \\
	\\
	\displaystyle \frac 1 {\rho_0} \: \partial_n \uinc(x) + \frac 1 {\rho_0} \:\partial_n u_0(x) &= 
	\frac 1 {\rho_1} \: \partial_n u_1(x)   \\
	\end{aligned}
	\right\} \; \text{for } x \in \Gamma_1\phantom{,} \\
	\\
	\left. \begin{aligned}
	u_1(x) &= \: u_2(x)  \\
	\\
	\displaystyle \frac 1 {\rho_1} \: \partial_n u_1(x) &= \frac 1 {\rho_2} \: \partial_n u_2(x)
	\end{aligned} \right\} \; \text{for } x \in \Gamma_2\phantom{,} 
	\label{trans_cond}
\end{array}
\end{equation}
where $\uinc(x)= e^{i {k}_0 \hat{k}_0 \cdot x}$ (an unitary incidence amplitude is assumed). 

Now, to recast this problem in a manner suitable to a BEM formulation (which is a method based on boundary integration),
the first step is to write the field $u$ in each region as linear combinations of certain kind of surface integrals
$ S, K $ called the single layer potential operator (SLP) and the double layer potential operator (DLP), respectively. 
An integral operator $U$ applied over a function $\varphi$ implies integration on a surface $\Gamma$ according to
\[
	U[ \varphi ](x) =  \int_\Gamma \Phi(x,y) \: \varphi(y) \: dS_y,
\]
where the kernel function $\Phi(x,y)$ gives the $x$-dependence (the integration variable is $y$).
Typically the $\varphi$-values on the surface $\Gamma$ will be the unknown quantities of an integral formulation for a 
boundary value problem. It should be noted that if $\Phi(x,y)$ satisfies Eq. \eqref{helmholtzpuro} with regarding the $x$-variable 
then $ U[\varphi](x)$ also does it, as long as the interchanging between the integral symnbol and the laplacian is possible. 
In the formulation presented here the $\Phi(x,y)$ is taken as the Green function or their derivatives, 
both choices satisfying Eq. \eqref{helmholtzpuro}. 
More precisely, in terms of integral operators the fields $u_i$ in each region are written as

\be
\begin{aligned}
 	u_0(x) &= \: d_{01} \: K_0 [\: \psi_1 \:](x) + s_{01} \: S_0[\:\phi_1\:](x) \quad &\text{for } x \in R_0 \\
	\\
	u_1(x) &= \: d_{11} \: K_1[\:\psi_1\:](x) + s_{11} \: S_1[\:\phi_1\:](x)  
	+ d_{12} \: K_1[\:\psi_2\:](x) + s_{12} \: S_1[\:\phi_2\:](x) \quad &\text{for } x \in R_1 \\ \label{fields_def}
	\\
	u_2(x) &= \: d_{22} \: K_2[\:\psi_2\:](x) + s_{22} \: S_2[\:\phi_2\:](x) \quad &\text{for } x \in R_2, 
\end{aligned}
\ee
where 
\be
\begin{gathered} 
	S_i[\:\phi_j\:](x) = \int_{\Gamma_j}G_{k_i}(x,y) \: \phi_j(y) dS_y \\
	K_i[\:\psi_j\:](x)= \int_{\Gamma_j}\partial_{n_y} G_{k_i}(x,y) \: \psi_j(y) dS_y,
	\label{slp_dlp_definitions}
\end{gathered} 	
\ee
are the SLP and the DLP operators, respectively, evaluated on the unknown functions $\phi_j, \psi_j$.
The subscript $j$ identifies the boundary $\Gamma_j$ ($j=1,2$) where the surface's integration is carried out.
The kernel $G_{k_i}(x,y)$ is the free-space 3D Green function for the Helmholtz equation in the wavenumber 
$k_i$, namely,
\[
	G_{k_i}(x,y) = \frac{e^{i k_i|x-y|}}{4\pi|x-y|},
\]
and the subscript $i$ identifies the region $R_i$ ($i=0,1,2$).
The coefficients $d_{ij},s_{ij}$ are real constants which will be fixed later. 


Notice that in regions $R_0$ and $R_2$ two operators (i.e. two degrees of freedom) are necessary to represent 
the related fields $u_0$ and $u_2$, whereas in the region $R_1$ four operators are required to represent $u_1$. 
This is because each field must fulfill two boundary conditions (b.c.) at each boundary that delimits its 
corresponding region. The fields $u_0$, $u_2$ must only verify the b.c. at $\Gamma_1$ and $\Gamma_2$, respectively, 
therefore two degree of freedom are sufficient in this case. On the contrary, the field $u_1$ must verify the b.c. 
simultaneously at both $\Gamma_1$ and $\Gamma_2$, requiring four degrees of freedom instead.


In order to build the normal velocity it is necessary to take the derivative with respect to the exterior normal of the
fields $u_i$, which are now expressed in terms of integral operators according to Eq. \eqref{fields_def}. 
This procedure leads to two new operators,
\ben
\begin{gathered} 
	K_i'[\: \phi_j  \:](x) = \partial_{n_x} \left( \int_{\Gamma_j}G_{k_i}(x,y)\phi_j(y) dS_y \right) \\
	T_i[\: \psi_j \:](x) = \partial_{n_x} \left(\int_{\Gamma_j} \partial_{n_y} G_{k_i}(x,y)\psi_j(y) dS_y\right),
\end{gathered}
\een
generically known as the \textit{normal derivative operators} \cite{Kress2001}.

The next step in the integral formulation of the problem is to evaluate the transmission conditions, Eq. \eqref{trans_cond},
using the field prescription according to Eq. \eqref{fields_def}. This process must be done considering the 
operator's {\it jump conditions} \cite{CK98} (i.e. its behavior in the limit when the evaluation point belongs to the 
integration surface). Then a system of four boundary integral equations for the four unknowns $\phi_j, \psi_j$ ($j=1,2$)
is obtained. By selecting the coefficients $\{ d_{ij},s_{ij} \}$ according to 
\begin{equation}
	\begin{array}{c}
	d_{01} = \rho_0, \quad d_{22} = \rho_2, \quad s_{01} = \rho_0^2, \quad s_{22} = \rho_2^2, \\
	d_{11} = \rho_1, \quad d_{12} = \rho_1, \quad s_{11} = \rho_1^2, \quad s_{12} = \rho_1^2,
	\end{array}
	\label{eleccionConstantes}
\end{equation}
the resulting boundary integral equation system is 
\begin{equation}
\small
\begin{aligned}
	\left. \begin{aligned}
	( \rho_0  K_0 - \rho_1  K_1 +\alpha_{01}) [\: \psi_1 \:](x) \: + \: (\rho_0^2  S_0 - \rho_1^2 S_1) [\: \phi_1 \:](x) \: + 
	\quad \qquad \qquad & \\
	- \rho_1 K_1[\: \psi_2 \:](x) - \rho_1^2 S_1[\: \phi_2 \:](x) = -\uinc(x) & \\
	\\
	-\muller_{01}[\: \psi_1 \:](x) + (\alpha_{01} - \rho_0 K'_0 + \rho_1 K'_1 )[\: \phi_1 \:](x) \: + \quad \qquad \qquad & \\
	+ \: T_1[\: \psi_2 \:](x) + \rho_1 K'_1[\: \phi_2 \:](x) = \frac{1}{\rho_0} \partial_n \uinc(x) & 
	\end{aligned} \right\} \text{ for } x \in \Gamma_1 \\
	\\
	\left. \begin{aligned}
	\rho_{1} K_{1}[\: \psi_{1} \:](x) + \rho_{1}^2  S_{1}[\: \phi_1 \:](x) + ( \rho_{1}  K_{1} - \rho_{2} K_2 + 
	\alpha_{12}) [\: \psi_2 \:](x) \: + \qquad  &\\
	+ \: ( \rho_{1}^2 S_{1} -\rho_{2}^2 S_2)[\: \phi_2 \:](x) = 0  & \\
	\\
	-T_{1}[\: \psi_{1} \:](x) - \rho_{1}K'_{1}[\: \phi_1 \:](x) - \muller_{12}[\: \psi_2 \:](x) + (\alpha_{12} -\rho_{1}K'_{1} + 
	\rho_{2}K'_2 ) [\: \phi_2 \:](x) = 0 & 
	\end{aligned} \right\} \text{ for } x \in \Gamma_2, \\
	\label{masterBIE}
\end{aligned}
\end{equation}
\normalsize
where $\alpha_{01}= (\rho_0 + \rho_1) /2$, $\alpha_{12}= (\rho_1 + \rho_2)/2$ and $\muller$ is
another operator, called the M\"uller operator and defined as
\be
	\muller_{s\ell}[\: \varphi \:](x) \equiv T_s [\: \varphi \:](x) -  T_\ell[\: \varphi \:](x).
	\label{Muller_operator}
\ee

The particular choice of constants in Eq. \eqref{eleccionConstantes} is aimed to force the turning up of the Müller operator,
Eq. \eqref{Muller_operator}, since its behavior regarding the singularity degree is advantageous over the use of the $T$ operator, 
which is hypersingular \cite{Kress2001}. 
Consequently, its numerical evaluation can be managed without special techniques (as mentioned by Ref. \cite{phdthesisMartin}
in section 5.2.1, for example). 

The system of boundary integral equations of Eq. \eqref{masterBIE} obtained in this section is based on Refs. \cite{kress1978transmission} 
and \cite{CK83}. The reader is referred to these references for a more detailed treatment.  

In summary, the solution of the acoustic scattering problem previously given in terms of the fields $u_i$ has been transformed 
in the search of the functions $\phi_j, \psi_j$ for each boundary $\Gamma_j$. These functions are called {\it densities}
in the literature and are now the unknowns. 

\subsection{Numerical method}

The system of Eq. \eqref{masterBIE} can be solved through a discretization process over the boundaries $\Gamma_j$,
which turns it in a finite-size matrix system. This leads to a BEM formulation.
For this step the standard procedure is to assume the following two approximations.
\begin{enumerate}
\item  Each surface $\Gamma_j$ is approximated by a planar triangular mesh (i.e., a set of triangles $\{ \Delta_\ell^j \} $ 
with $\ell = 1,2,.. N_j$), so that 
\ben
	\Gamma_j = \bigcup_{\ell=1}^{N_{j}} \Delta_\ell^j,
\een
where $\Delta_\ell^j$ is the $\ell$-th triangle whose centroid is $x_\ell^j$ and $N_j$ is the total number of triangles
of the mesh that represents the $\Gamma_j$ boundary.
\item
The unknown densities $\psi_j$ and $\phi_j$ are considered as piecewise constant functions in each triangle, that is,
\be
	\psi_j(x) = \sum_{\ell=1}^{N_j} \psi^j_\ell  I_{\Delta_\ell}(x) \qquad
	\phi_j(x)= \sum_{ \ell=1}^{N_j} \phi^j_\ell  I_{\Delta_\ell}(x),
	\label{discr_fields}
\ee
where $\psi^j_\ell, \phi^j_\ell$ are unknown complex numbers and $ I_{\Delta_{\ell}}(x) $ is the indicator function of the 
$\ell$-th triangle, defined as 
\[
	I_{\Delta_{\ell}}(x) = \begin{cases}
	                        1 \quad \text{ if } x \in \Delta_\ell \\
	                        0 \quad \text{ otherwise}
	                       \end{cases}
\]
\end{enumerate} 

In order to find the densities, the prescription given in Eq. \eqref{discr_fields} is introduced in the system of Eq.
\eqref{masterBIE}. 
This procedure transforms each integral over the boundary $\Gamma_j$ into a sum of integrals over each triangle 
$\Delta_\ell^j$.

The resulting system remains, of course, valid for all $x \in \Gamma_j$, so that in particular 
is valid for the set of centroids $\{x_\ell^j\} (\ell=1,2,..,N_j)$ belonging to the boundary $\Gamma_j$. 
When these discretized equations are evaluated in both sets $( \{x_\ell^1\}, \{x_\ell^2\} )$, a matrix system of size $m \times m$ 
is obtained, with $m= 2(N_1 + N_2)$. The unknown values are the complex quantities $\{\psi^j_\ell, \phi^j_\ell \}$.

These equations can be expressed as a square matrix system, namely,
\begin{equation}
	\boldsymbol{A}
	\begin{pmatrix}	\boldsymbol{ \psi^1 } \\
	\boldsymbol{ \phi^1 } \\
	\boldsymbol{ \psi^2 } \\
	\boldsymbol{ \phi^2 }
	\end{pmatrix} =
	\begin{pmatrix}
	\boldsymbol{ f } \\
	\boldsymbol{ g }\\
	\boldsymbol{ 0 } \\
	\boldsymbol{ 0 }
	\end{pmatrix} 
	\label{sistema}
\end{equation}
where the square matrix $\boldsymbol{ A }$ has $2(N_1+N_2)$ rows, and the unknowns $\boldsymbol{ \psi }^j, 
\boldsymbol{ \phi}^j $ and data vectors $\boldsymbol{ f }, \boldsymbol{ g }$ are defined as 
\[
	\boldsymbol{ \psi^j } = \begin{pmatrix}
				\psi^j_1 \\
				\\
				\psi^j_2 \\
				\\
				\vdots \\
				\\
				\psi^j_{N_j}
				\end{pmatrix}
	\qquad
	\boldsymbol{ \phi^j } = \begin{pmatrix}
				\phi^j_1 \\
				\\
				\phi^j_2 \\
				\\
				\vdots \\
				\\
				\phi^j_{N_j}
				\end{pmatrix}
	\qquad
\]
\begin{equation}
	\boldsymbol{ f } = -\begin{pmatrix}
				\uinc(x_1^1) \\
				\\
				\uinc(x_2^1) \\
				\\
				\vdots \\
				\\
				\uinc(x_{N_1}^1)
				\end{pmatrix}
	\quad 		
	\boldsymbol{ g } = \frac{1}{\rho_0} \begin{pmatrix}
				\partial_n\uinc(x_1^1) \\
				\\
				\partial_n\uinc(x_2^1) \\
				\\
				\vdots \\
				\\
				\partial_n\uinc(x_{N_1}^1)
				\end{pmatrix}.
	\quad 	
	\label{fgdef}	
\end{equation} 
The $\boldsymbol{A}$-matrix full expression is  given in the next subsection.

\subsubsection{Matrix definition}
\label{app_matrix_operators}

The matrix system has a symmetry which is emphasized in the four-block structure of submatrices $\boldsymbol{B}_i$, 
$\boldsymbol{D}_i$ and $\boldsymbol{I_i}$ ($i=1, 2$), namely,
\be
	\boldsymbol{A}=	\begin{pmatrix}
	\boldsymbol{D}_1 & \boldsymbol{B}_1 \\
	\boldsymbol{B}_2 & \boldsymbol{D}_2
	\end{pmatrix} +
	\begin{pmatrix}
	\alpha_{01} \boldsymbol{I}_1 & \boldsymbol{0} \\
	\boldsymbol{0} & \alpha_{12} \boldsymbol{I}_2
	\end{pmatrix}. 
	\label{system_matrix}
\ee
The matrices $\boldsymbol{I}_1$ and $\boldsymbol{I}_2$ are identities with dimensions $2N_1$ and $2N_2$, respectively. 
Each $\boldsymbol{D}_i $  is a matrix of $2N_i \times 2N_i$ size and has the form
\begin{equation}
	\boldsymbol{D_1} = 
	\begin{pmatrix}
	 ( \rho_0\boldsymbol{K}_0 - \rho_1\boldsymbol{K}_1 )^{[1,1]} & 
		(\rho_0^2\boldsymbol{S}_0-\rho_1^2 \boldsymbol{S}_1)^{[1,1]} \\
		\\
		-\boldsymbol\muller_{01}^{[1,1]} & 
		( -\rho_0 \boldsymbol{K'}_0 + \rho_1 \boldsymbol{K}'_1 )^{[1,1]}
	\end{pmatrix}
	\label{matrizD1}
\end{equation} 
\[
	\boldsymbol{D_2}= 
	\begin{pmatrix}
	( \rho_{1}  \boldsymbol{K}_{1} - \rho_{2}  \boldsymbol{K}_2  )^{[2,2]} &
	(\rho_{1}^2 \boldsymbol S_{1} - \rho_{2}^2 \boldsymbol S_2)^{[2,2]} \\
	\\
	{-\boldsymbol \muller_{12}}^{[2,2]} & 
	( -\rho_{1}\boldsymbol K'_{1} + \rho_{2}\boldsymbol K'_2 )^{[2,2]} \\
	\end{pmatrix}.
\]
The matrices $\boldsymbol{B}_1, \boldsymbol{B}_2$ have size $2N_1 \times 2N_2$ and $2N_2 \times 2N_1$, respectively,
and its expressions are
\[
	\boldsymbol{B_1} = 
	\begin{pmatrix}
	-\rho_1\boldsymbol{K}_1^{[1,2]} & - \rho_1^2  \boldsymbol{S}_1^{[1,2]}\\
	\\ 
	\boldsymbol{T}_1^{[1,2]}  & \rho_1 {\boldsymbol K'}_1^{[1,2]} \\
	\end{pmatrix},
\]

\[
	\boldsymbol{B_2}= 
	\begin{pmatrix}
	\rho_{1} \boldsymbol{K}_{1}^{[2,1]} & \rho_{1}^2 \boldsymbol{S}_{1}^{[2,1]} \\
	\\
	-\boldsymbol T_{1}^{[2,1]}& -\rho_{1}{\boldsymbol K'_{1}}^{[2,1]}
	\end{pmatrix}.
\]
The discrete version of a generic operator $\boldsymbol{U}^{[a,b]}_q \in \mathbb{C}^{N_{a} \times N_{b}}$ 
($a,b=1,2$) with kernel $\Phi(k_q ; x, y )$ follows the notation 
\[
	(\boldsymbol{U}^{[a,b]}_q )_{\ell s}= \int_{\Delta_s^b} \Phi(k_q; x_\ell^a,y) dS_y
\]
for the $\ell s-$element. Thus, the first superindex ($a$) refers to the boundary where the evaluation point is located whereas the second ($b$)
refers to the boundary to which  the triangle $\Delta^b$ over which the surface integration is carried out belongs.
The matrix row-index $\ell$ is associated with a particular evaluation point $x_\ell$ while the column index $s$ 
is associated with the particular element $\Delta_s$.
The subindex $q$ identifies the corresponding wavenumber $k_q$.
For example, 
\[
	\left(\boldsymbol{S}_{1}^{[1,2]}\right)_{\ell s} =\int_{\Delta_s^2} G_{k_1}(x_\ell^1,y) \: dS_y,
\]
implies integration over the $s$-th triangle of the boundary $\Gamma_2$ and evaluation on centroid $x_\ell^1$ belonging 
to the boundary $\Gamma_1$, all for the wavenumber $k_1$

\subsection{Scattered field and TS computation}

Once the densities $\{ \psi_j,\phi_j \}$ have been obtained, the scattered field at an exterior point $x$ 
can be calculated by evaluating the first Eq. in \eqref{fields_def} with the piecewise approximation made at 
Eq. \eqref{discr_fields}. 
Therefore, the discretized version of the external field is 
\be
	u_0(x) = \: \sum_{ \ell=1}^{N_1}  \left( \rho_0 \: \psi^1_\ell \int_{\Delta_\ell} \partial_{n_y} G_{k_0}(x,y) \: dS_y \: + 
	\rho_0^2 \: \phi^1_\ell \int_{\Delta_\ell} G_{k_0}(x,y) \: dS_y \right).
	\label{nearfieldEq}
\ee

When the exterior point $x$ is located far away the scatterer object ($|x| \to \infty$), 
it is possible to use the asymptotic expression of the Green function and its normal derivative, namely,
\[
	G_{k_0}(x,y) = \frac{\euler{ik_0|x|}}{ 4 \pi |x|} \left[ \euler{-ik_0\hat{x}\cdot y} + 
	\mathcal{O}\left(\frac{1}{|x|}\right)\right]
\]
\[
	\partial_{n_y} G_{k_0}(x,y) = \frac{\euler{ik_0|x|}}{ 4 \pi |x|} \left[\frac{\partial 
	\euler{-ik_0\hat{x}\cdot y }}{\partial n_{y}} + \mathcal{O}\left(\frac{1}{|x|}\right)\right],
\]
where $\mathcal{O}(f(x))$ means that there exists a positive constant $C$ such that $\mathcal{O}(f(x))\leq C f(x)$, 
thus, if $f(x)$ goes to zero when $|x| \to \infty$, then $\mathcal{O}(f(x))$ also does it.

Subtituting the above expressions into Eq. \eqref{nearfieldEq} the scattered field turns out 
\[
	u_{0}(x) = \frac{e^{ i k_0|x|}}{|x|} \left[ f_\infty(\hat{x}) + \mathcal{O}\left( \frac{1}{|x|} \right) \right],
\]
where $\hat{x}= x / |x|$ is the unit vector in the direction of observation --pointing towards the observer--, $k_0$ is 
the wavenumber of the incident field and $f_\infty$ is the farfield scattering amplitude (a quantity with length's units)
whose expression is
\be
	f_\infty(\hat{x}) = \frac{1}{4\pi} \sum_{\ell=1}^{N_1} \left( 
	-i k_0 \psi_\ell^1 \rho_0 \int_{\Delta_\ell} \euler{- ik_0\hat{x}\cdot y }
	\hat{x} \cdot n_{y} \: dS_y \; +
	\rho_0^2 \phi^1_\ell \int_{\Delta_\ell} \euler{-ik_0\hat{x}\cdot y } \: dS_y 
	\right).
	\label{finf_bs_discretized} 
\ee

Thus, the usual cases of back-scattering and forward-scattering are obtained by considering $\hat{x} = -\hat{k}_0 $ and    
$\hat{x} = \hat{k}_0 $, respectively, i.e.
\be
	f_{\infty}^{\text{bs}} \equiv f_\infty(\hat{x} = -\hat{k}_0 ) \qquad  \qquad
	f_{\infty}^{\text{fw}} \equiv f_\infty( \hat{x} = \hat{k}_0 ).
	\label{convencion_back_forward}
\ee

In fisheries acoustics, as well as in other SONAR applications, sound scattering by an object is analyzed in 
the logarithmic scale using the target strength  (TS) parameter that can be expressed as 
\be
	\text{TS} = 10 \log_{10} \left( |f_\infty|^2 \right) \text{ dB re 1 m$^2$}.
	\label{TS_def}
\ee

As a summary, once $\boldsymbol{\psi}^1$ and $\boldsymbol{\phi}^1$ from Eq. \eqref{sistema} are known, the TS parameter 
can be computed by using the numerical evaluation of Eq. \eqref{finf_bs_discretized} in the formula given by Eq. \eqref{TS_def}.

\section{Model verifications}
\label{verifications}

In order to verify the formulated model, identified from now on as the Coupled BEM model, two types of tests are 
conducted, by comparisons between model predictions and benchmark results derived from exact solutions.

In first place, the acoustic scattering problem corresponding to a single penetrable obstacle, a prolate spheroid,
is considered. Consequently only a scattering boundary $\Gamma$ is present and the formulation of the problem is simplified 
(a brief description of this case is worked out in the Appendix).
In second place, the acoustic problem of two concentric spheres is considered. This allows to explicitly test the system 
of Eq. \eqref{sistema}.

\subsection{Single fluid acoustic problem}
\label{valid_fluid}

For a single fluid obstacle, Coupled BEM model predictions are compared
with a benchmark solution previously reported \cite{jech2015comparisons}. 
These authors compute backscattering TS as a function of the incidence angle $\theta$ for a gas-filled prolate spheroid 
with semi-axis $ a = 0.07 $ m and $ b = 0.01 $ m, at $ 38 $ kHz.
In order to compute the scattering response of the same prolate spheroid using the model, a spheroidal mesh with $N = 44480$ triangular 
elements is used.
The number of elements in the mesh is selected in order to guarantee that the acoustic wavelength would be several times greater than 
the distance between vertices (usually five or six times greater) following the recommendation reported in \cite{francis2003depth}.

\begin{figure}[hbt]
	\centering
	\includegraphics[scale=0.4]{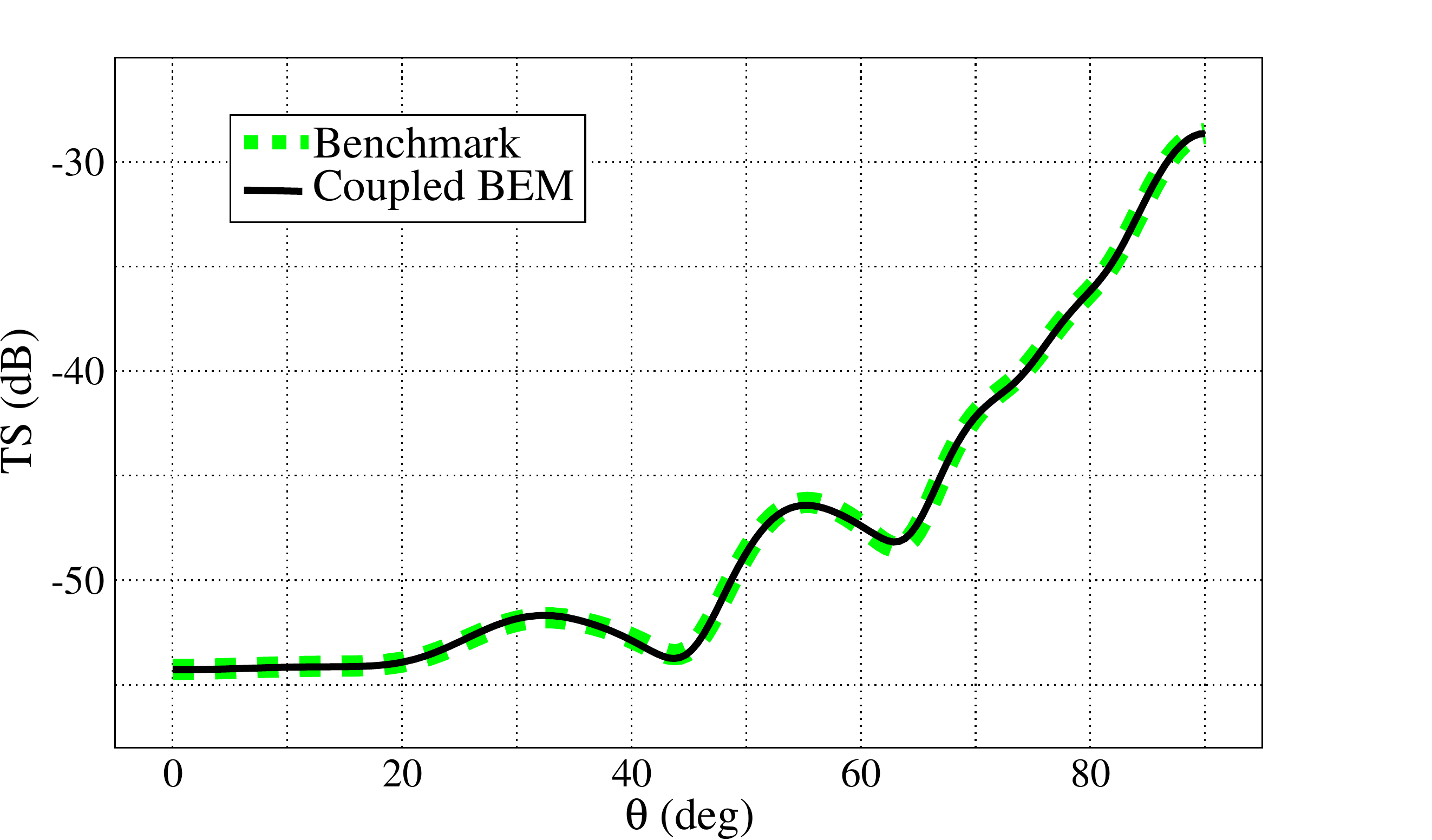}
	\caption{Comparison of TS \textit{vs.} incidence angle $\theta$ for a gas-filled prolate spheroid with aspect ratio 1:7, 
	evaluated according to the Coupled BEM model (solid line) and to the benchmark solution (dotted line).
	 }
	\label{fig_bench_sph_gas_filled} 
\end{figure}

The comparison between the modelled TS and the TS computed for the exact prolate spheroid solution \cite{gonzalez2016computational},
is exhibited in Figure \ref{fig_bench_sph_gas_filled}, where a good agreement is observed. 
The assumed sound speed and density for the gas inside the prolate spheroid were $ c_1 = 345.0 $ m s$^{-1}$ and
$ \rho_1 = 1.24 $ kg m$^{-3}$, respectively; whereas the corresponding values in the surrounding medium (water) were 
$c_0 = 1477.4$ m s$^{-1}$ and $\rho_0 =1026.8$ kg m$^{-3}$. These values were taken from the literature (see Table II from 
Ref. \cite{jech2015comparisons}) and corresponds to realistic values in aquatic ecosystem research applications.

\subsection{Two fluids acoustic problem}
\label{valid_fluid_two_fluid}

The acoustic scattering problem for two 3D bounded fluid objects has no exact analytical solution, except in the case of two concentric spheres.
In this case, the solution is given in terms of an analytical modal series (partial wave decomposition) coming from the 
separation of the wave equation in spherical coordinates \cite{hickling1964analysis}.

In order to compare the values predicted by the Coupled BEM model against the benchmark modal solution,
two concentric spheres of radii $ r_1 = 0.06 $ m and $ r_2 = 0.016 $ m are considered,
where the subscript $1$ refers to the external sphere and the subscript $2$ to the internal one.
Keeping in mind, as in the previous example, applications to fisheries acoustics, the values of sound speed $c$ and density 
$\rho$ of the material media are taken from \cite{gorska2003modelling}, and they are also listed in Table \ref{Tabla_datos}.
The medium ``0'' corresponds to the surrounding water, while media ``1'' and ``2'' correspond to the external and internal 
spheres, respectively.

\begin{table}[ht]
	\centering
	\begin{tabular}{ccc}
	\hline\hline
	Medium & $c$ (m s$^{-1}$) & $\rho$ (kg m$^{-3}$) \\
	\hline
	0 (water) & 1477.4 & 1026.8 \\
	1 & 1.04 $c_0$ & 1.04 $\rho_0$ \\
	2 & 0.23 $c_0$ & 0.00129 $\rho_0$ \\
	\hline\hline
	\end{tabular}
	\caption{Material properties (sound speed $c$ and density $\rho$) for the media 0, 1, 2
	in the two spheres acoustic scattering problem.}
	\label{Tabla_datos}
\end{table}

\begin{figure}[htb]
	\centering
	\includegraphics[scale=0.4]{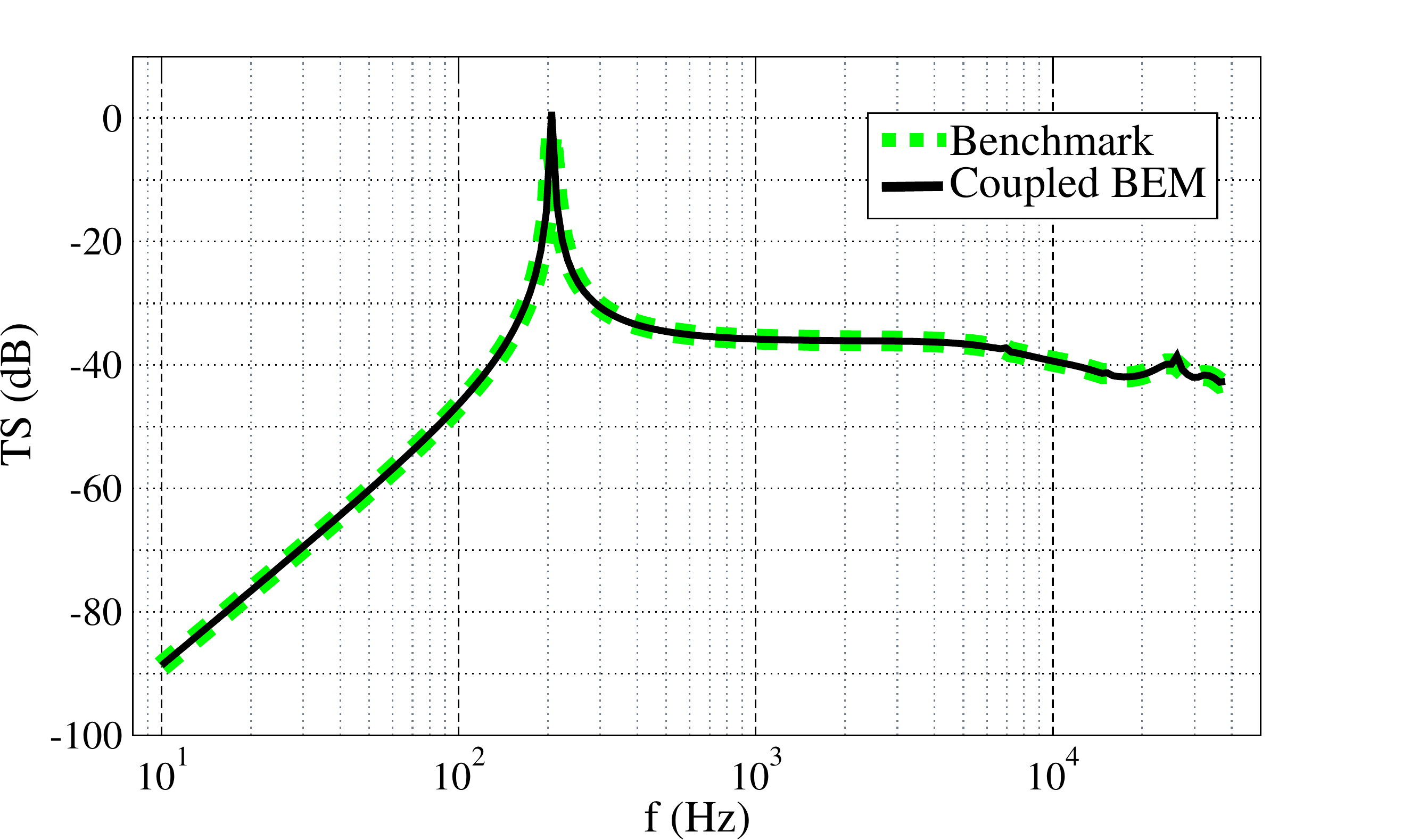}
	\caption{ Comparison of TS \textit{vs.} frequency for two concentric spheres evaluated according to the Coupled BEM model (solid line) and 
	to the benchmark solution computed through the modal series (dotted line).}
	\label{fig_bench_spheres} 
\end{figure}

For modelling this problem under the Coupled BEM approach, two spherical meshes were built. The number of triangular elements were  
$ N_1 \!=\! 1142 $  and $ N_2\! =\! 2274 $, for the spheres of radii $r_1$ and $r_2$ respectively. 
Backscattering TS values for the frequency range $0.01 - 38$ kHz, were computed using the modal series and the BEM implementation.
Results of their comparison are shown in Figure \ref{fig_bench_spheres}, where a good agreement is again evident. 

\subsection{Error in TS computations}

The criterion reported in Ref. \cite{jech2015comparisons} is used to estimate the error level of the modelled TS in the previous examples.
In that work the authors quantified the error as the mean of the absolute deviation between the modelled TS and the TS computed in the 
benchmark case (exact solution), i.e.
\[
	| \overline{\Delta \TS} | = 
	\frac{1}{L}\sum_{i=1}^L |\TS_i(\mbox{\scriptsize prediction})-\TS_i(\mbox{\scriptsize benchmark})|,
\]
where $L$ is the total number of frequency or angle calculated values.
Values for $|\overline{\Delta \TS}|$ were $ 0.066 $ dB and $ 0.12 $ dB for the prolate spheroid and the two concentric spheres,
respectively.

\section{Application to fish tomography}
\label{application}

The {\it Merluccius hubbsi} is one of the most important fishing resources of the Argentine Sea. 
Moreover, it is the most abundant species of  demersal fish (living near the seabed) in the southwestern Atlantic Ocean. 
Therefore, it plays a prominent role in the marine ecosystems of the Argentinian continental shelf.
It is also important to note that their extraction crucially depends  on the possibility of detection of fish schools using 
underwater acoustics. 
For this reason, a better knowledge of the backscattering of this species, as well as its acoustic scattering in other directions
is very important both for fishing activity and  for scientific research purposes.

\subsection{ Meshes and scattering parameters}

Computerized tomography (CT) scans performed on a {\it Merluccius hubbsi} specimen, with spatial resolution of 1.5 mm, allowed
for building two 3D triangular meshes to represent the fish body and its swimbladder having $N \! =\! 8983$ and $N\!=\!11474$ 
elements, respectively.

\begin{figure}[!htb]
	\centering
	\includegraphics[scale=0.4]{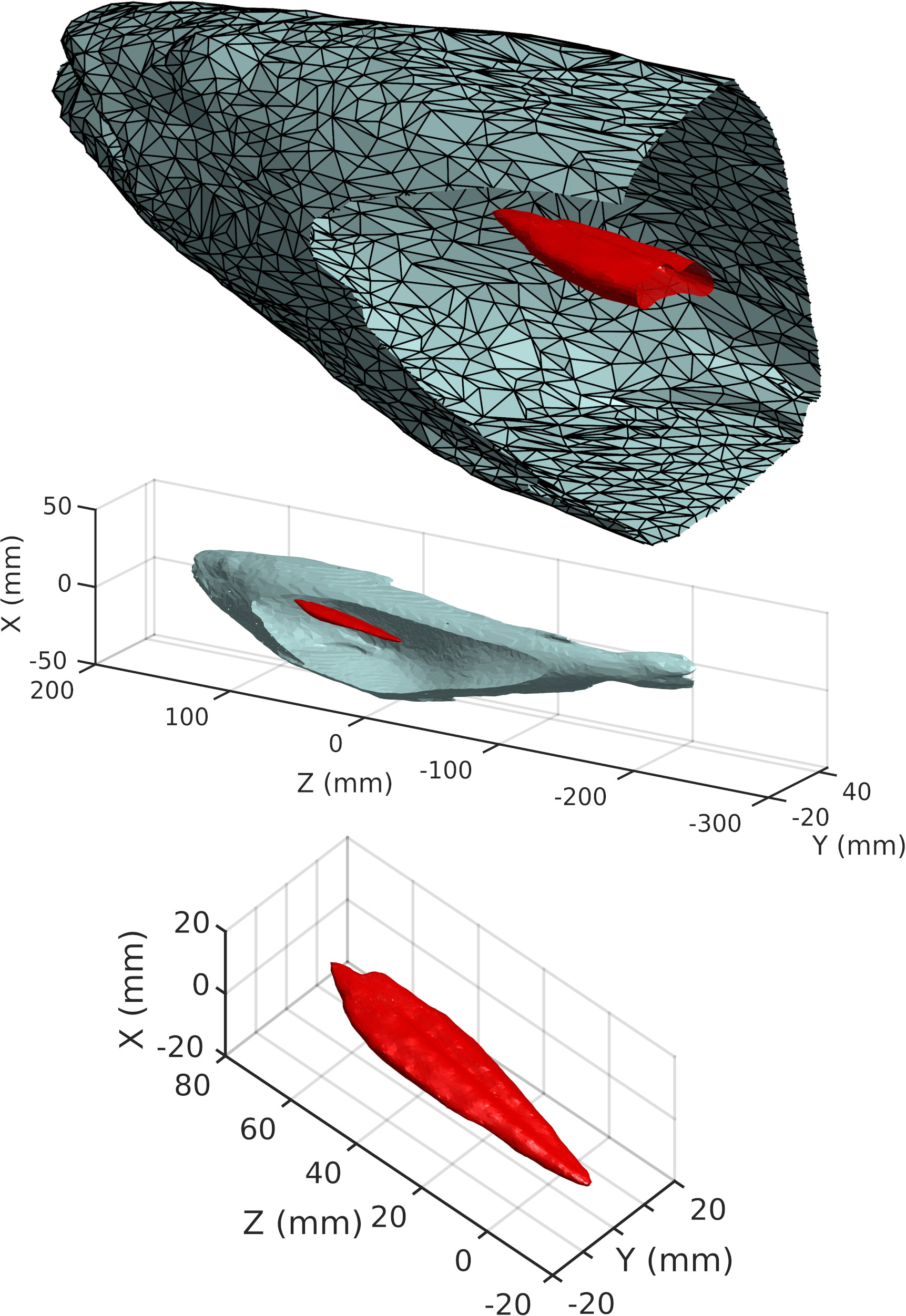}
	\caption{3D meshes generated from Computer Tomography scan.The top panel shows
	a detailed view of the fish's head where individual triangular facets are appreciable. The middle panel displays the body-mesh
	and the swimbladder-mesh inside it so that their relative locations can be visualized. A more detailed view of the swimbladder
	mesh is shown in the bottom panel, where for clarifying purposes the view has been rotated.
	}
	\label{fig_mesh_body} 
\end{figure}

Meshes are exhibited in Figure \ref{fig_mesh_body}. 
Top panel shows the fish head and part of the swimbladder. Triangle edges that constitute the fish body-mesh are visualized. 
In the middle panel both scatterers are exhibited and it can be noticed that their relative locations are not concentric. Moreover,
the swimbladder longitudinal axis is approximately  $10^\circ$ tilted head-up respect to the main body axis.
The swimbladder is presented in detail in the bottom panel, its  geometry is rather complex and cannot be strictly represented
by a simple shape such as a sphere, a finite-length cylinder or a spheroid.


The material properties of the surrounding medium (water),
the fish flesh constituting the body and the gas in the swimbladder are taken from the Table \ref{Tabla_datos}.
All the simulations were computed at $f=38$ kHz, since it is a usual frequency in fisheries acoustics. 
The body length is 382.5 mm and the scattering can be characterized by the dimensionless parameter 
$k_0 a \approx 31 $. For the swimbladder, whose length is 81.8 mm, there are two relevant parameters, namely,
$ k_1 a \approx 6.3 $ and $ k_2 a \approx 28.7 $. In both cases $a$ is the appropriate semi-longitude of the 
scatterer.

\subsection{Backscattering and forward scattering TS computations}

\begin{figure}
	\centering
	\includegraphics[scale=0.5]{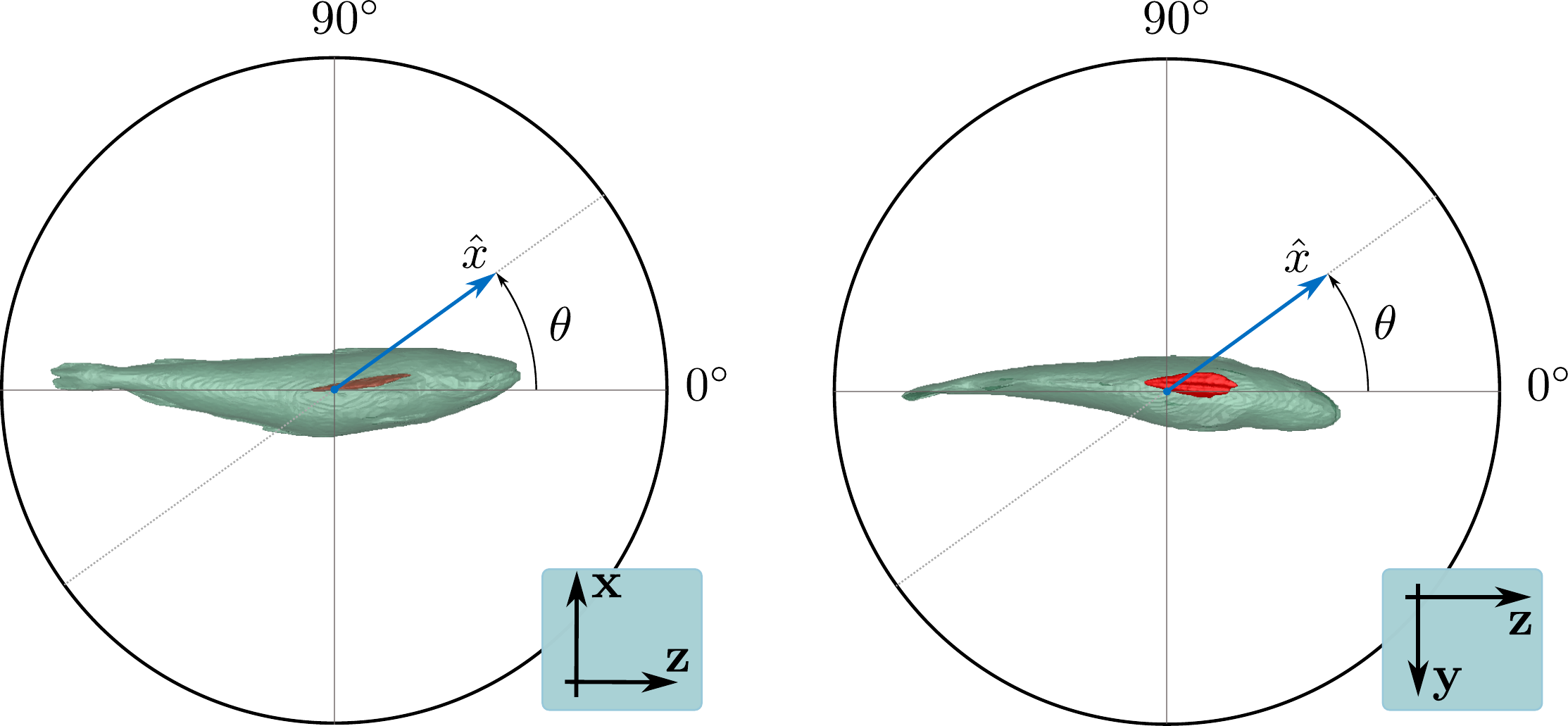}
	\caption{ Schematic grid for evaluation of TS by fish in the {dorsal-ventral} aspect (left) and lateral 
	aspect (right). The angle $\theta$ is associated to the observation direction $\hat{x}$ as it is illustrated in both panels. 
	To emphasize the swimbladder relative location respect to the fish body, the latter is shown with transparency.}
	\label{fig_scheme_ff} 
\end{figure}

Since at 38 kHz both scatterers (body and swimbladder) have an acoustic length of several wavelengths it is expected
that the scattering, mainly the backscattering, strongly depends upon morphology and orientation.
Keeping these considerations in mind, backscattering and forward scattering evaluations as a function of the 
observation angle $ \theta $ are conducted for dorsal-ventral aspect (incidence contained in the $ y = 0 $ plane, 
see scheme in Figure \ref{fig_scheme_ff} left) and lateral aspect (incidence contained in the $ x = 0 $ plane, see Figure 
\ref{fig_scheme_ff}, right), for the entire rotation $0 \leq \theta \leq 360^\circ$ in both cases.

The observation angle $ \theta $ corresponds to the observation direction $\hat{x}$, as it is stated in the Figure \ref{fig_scheme_ff}.
Thus, according to Eq. \eqref{convencion_back_forward}, for the backscattering case the incidence direction $\hat{k}_0$ is opposite 
to $\hat{x}$ while in the forward scattering both directions are coincident, namely $\hat{k}_0 = \hat{x}$.

\subsubsection{Backscattering computations}

The resulting patterns for backscattering TS are presented in this section. As stated in the Section \ref{Intro}, other investigators 
\cite{francis2003depth,reeder2004broadbandExperimento} have modelled the TS of a fish considering only the swimbladder and neglecting 
the body contribution. This formulation implies 
\be
	\TS = 10 \log_{10}( |f_{\infty,Sb}^{\text{bs}}|^2 ),
	\label{TSswim}
\ee
where $f_{\infty,Sb}^{\text{bs}}$ is the farfield backscattering yielded by the swimbladder. 
On the other hand, TS has been also modelled by computing both contributions separately and then adding them 
incoherently \cite{gorska2003modelling} or coherently \cite{prario2015prolate,reeder2004broadbandExperimento}. 
In particular, the coherent addition leads to
\be
	\TS = 10\log_{10}(|f_{\infty,Sb}^{\text{bs}} + f_{\infty,B}^{\text{bs}}|^2), 
	\label{TSswimBody}
\ee
where $f_{\infty,B}^{\text{bs}}$ is the corresponding farfield pattern exclusively from the body.
It should be emphasized that a formulation like the proposed in \eqref{TSswimBody} is an approximation because each $f_{\infty}$ is
calculated {\it independently}, i.e, without interaction between the scatterers.
The BEM approach presented in this work treats the problem including both contributions jointly, within a rigorous framework for
a coupled system.

In order to compare the results of the Coupled BEM approach with the aforementioned approximated models, it is useful to establish the 
following nomenclature: All the TS calculated by using the Coupled BEM approach, which is derived from Eq. \eqref{finf_bs_discretized}, 
will be labelled ``Model'';
TS patterns obtained considering only the acoustic response from the swimbladder, derived from Eq. \eqref{TSswim} and computed through the 
single scatterer BEM approach (see Appendix) will be labelled ``Sb'' and, finally, patterns obtained for the coherent 
sum of the body and swimbladder acoustic responses, according to Eq. \eqref{TSswimBody} and calculated through the evaluation of each 
$f_\infty$ separately, both using the single scatter BEM approach, will be labelled ``Sb + B''.

\begin{figure}[!h]
	\centering
	\includegraphics[scale=0.75]{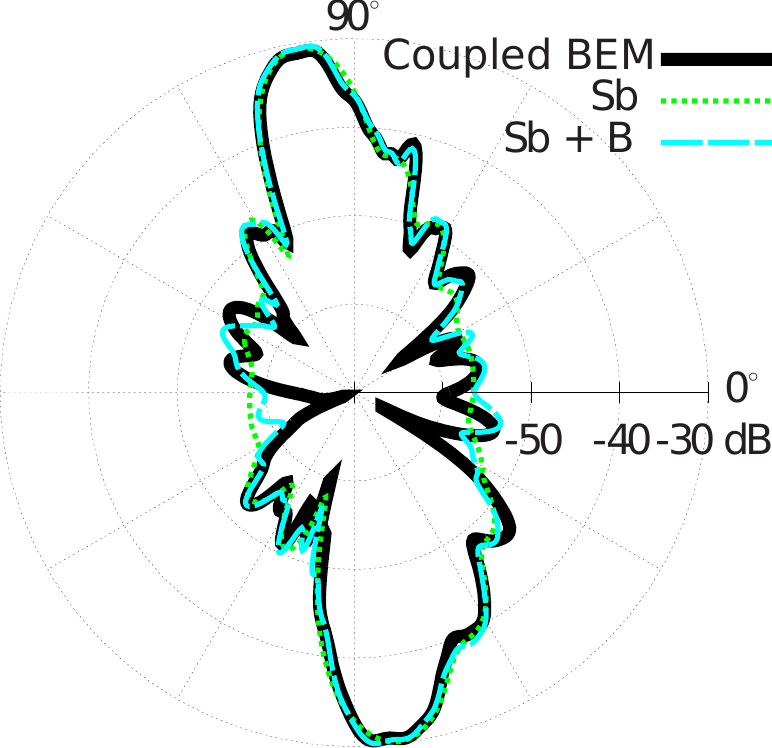}
	\hskip 3em
	\includegraphics[scale=0.75]{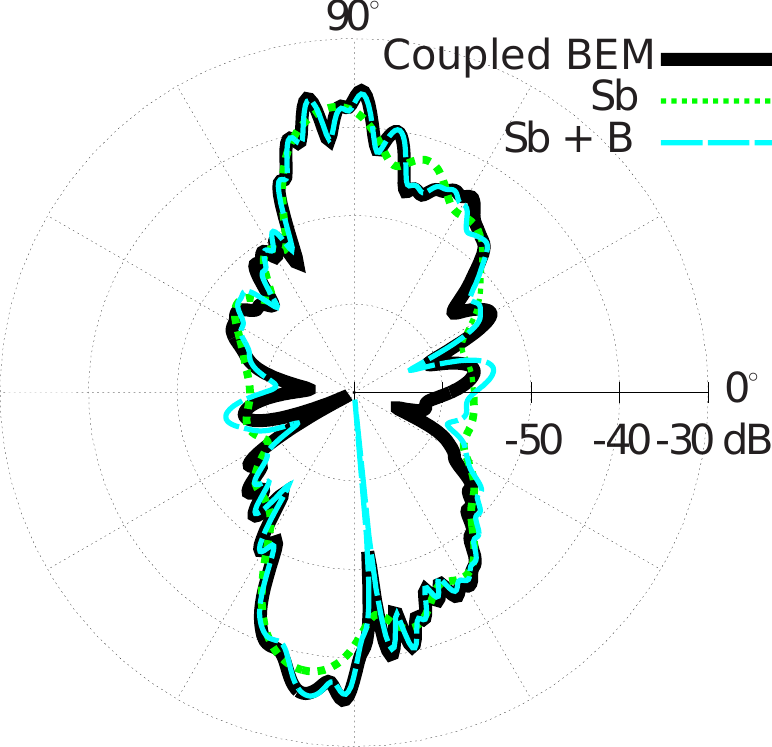}
	\caption{Backscattering TS as a function of angle $ \theta $ in the dorsal-{ventral} aspect (left panel) and lateral 
	aspect (right panel). Swimbladder and fish body are considered as a gas-filled and weakly scattering 
	objects, respectively. Both contributions are taking into account in different ways. In the Coupled BEM approach (solid 
	line) the interaction between swimbladder and fish body is explicitely considered. The curve ``Sb'' shows exclusively 
	the response of the swimbladder (dotted line) while the curve ``Sb+B'' shows the response of the swimbladder coherently 
	added to the fish body one (dashed line).}
	\label{fig_polar_bs}
\end{figure}

The resulting patterns for backscattering TS according to the three types of evaluation are shown in Figure \ref{fig_polar_bs} 
as a polar plot. In the left panel (dorsal-ventral aspect) it is clear that the presence of TS maxima at 99$^\circ$ and 277$^\circ$ is associated with
orientations where the swimbladder yields a greater normal surface (see Figure \ref{fig_scheme_ff}, left) with respect to the
incidence direction.
Those orientations are precisely the directions where the swimbladder contribution and the fish body plus the swimbladder
as a whole, shows the best match.
However, that match deteriorates near the longitudinal axis of the fish, i. e. in the head-tail direction, where the interaction
between body and swimbladder, explicitly considered by the coupled BEM approach, is relevant and therefore the coherent sum does 
not provide a good match of the corresponding curves.

When the lateral aspect of the fish is considered (Figure \ref{fig_polar_bs}, right panel), it is found that the curve that exclusively 
represents the contribution of the swimbladder differs from the curve generated by the model to a great extent in comparison with the 
previous case (i.e. dorsal-ventral aspect). 
This observation can be justified because the insonified area that the swimbladder offers to the acoustic incident wave is undoubtedly
smaller at the fish lateral aspect than at the fish dorsal-ventral one. 

\begin{figure}[!hbt]
	\centering
	\includegraphics[scale=0.12]{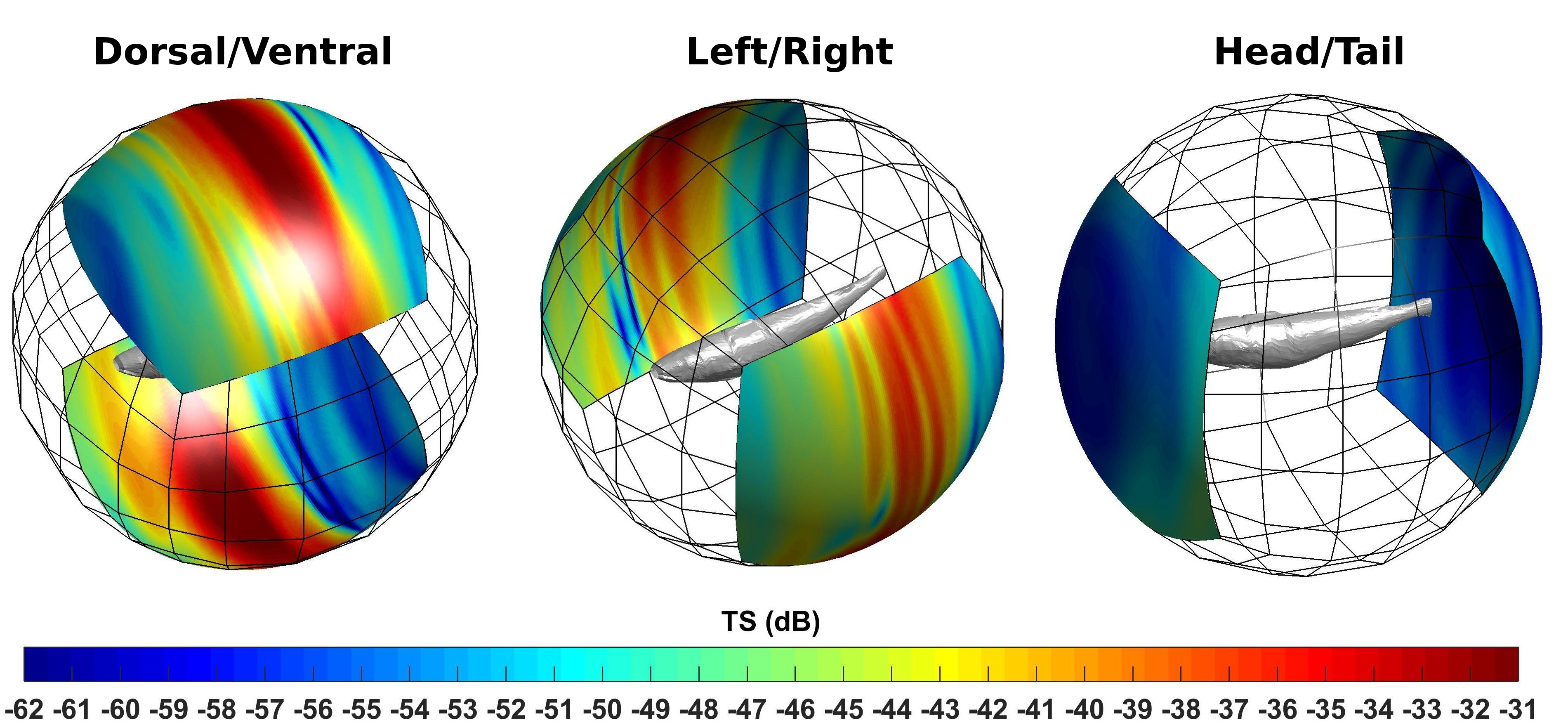}\\
	\caption{ 
	3D Vizualization of backscattering TS by fish for incident angles represented on the surface of a sphere. 
	Each direction is identified with a point on the surface of a sphere surrounding the scatterer and its colour provides a 
	measure of the backscattering TS magnitude. The sphere is divided into six patches that are shown in pairs Dorsal/Ventral
	(left panel), Left/Right (middle panel) and Head/Tail (right panel). Within each sphere the mesh of the fish is plotted 
	in order to indicate its relative location.
	}
	\label{fig_planeta}
\end{figure}

Model results of the TS parameter are traditionally presented as 2D polar plots as shown in Figure \ref{fig_polar_bs}.
A proper 3D plot visualization of the scattering phenomenon has been reported (e.g. see  \cite{towler2003visualizing}).
An alternative type of visualization for backscattering responses is proposed in this work for all possible incidence $\hat{k}$ directions, 
as it can be observed in Figure \ref{fig_planeta}, where each direction is identified with a point on the surface of a sphere surrounding 
the scatterer and its colour provides a measure of the backscattering TS magnitude. With the purpose of helping the visualization of the 
results in the corresponding plots, the sphere has been divided into six patches which are shown in pairs in Figure \ref{fig_planeta}.
These pairs of patches are identified with the names that appear above them (i.e. Dorsal/Ventral, Left/Right and Head/Tail). 
Additionally, within each sphere the mesh of the fish has been plotted in order to indicate its relative location. 
For example, maximum TS is achieved at the directions whose colour has the highest value in the colour-bar scale located
at the bottom side of the figure. That happens when swimbladder is approximately orthogonal to the incident wave front. 
The ranges of TS values obtained for each pair of patches were $[-82.1, -29.9]$ dB, $[-89.02 -32.1]$ dB and $[-63.1 -45.1]$ dB, 
respectively.

\subsubsection{Forward scattering computations}

Although forward scattering has been studied less extensively than backscattering (partly because of its greater experimental difficulty) it also has interesting applications. The forward scattering can be applied in fish detection methodologies \cite{diachok2000absorption}, therefore it has been addressed 
with simplified models \cite{ding1997method}, \cite{ye1996acoustic}. 
Furthermore, the estimation of attenuation through the medium is related to the forward scattering $f_\infty$ \cite{chu1999phase}.

\begin{figure}[!hbt]
	\centering
	\includegraphics[scale=0.75]{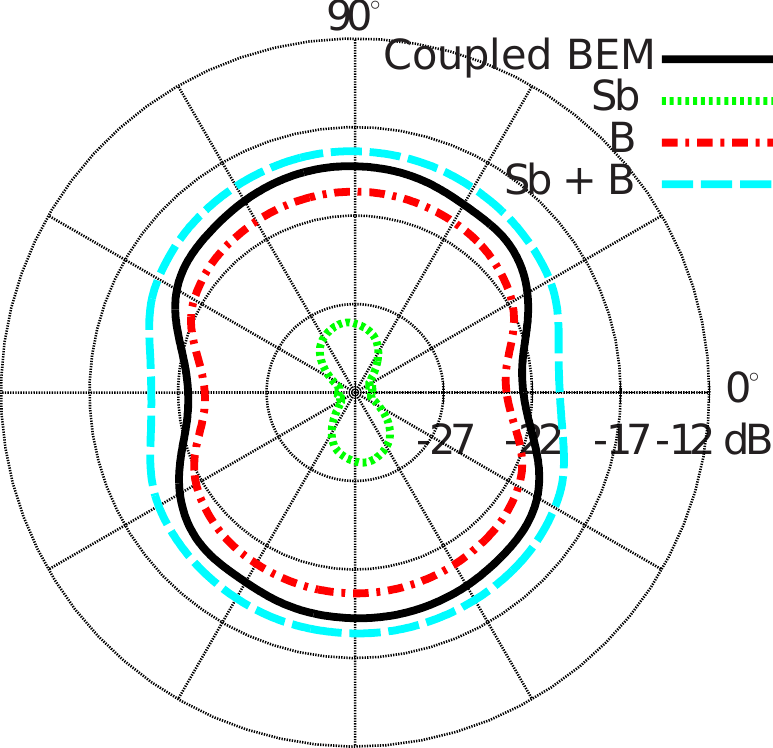}
	\hskip 3em
	\includegraphics[scale=0.75]{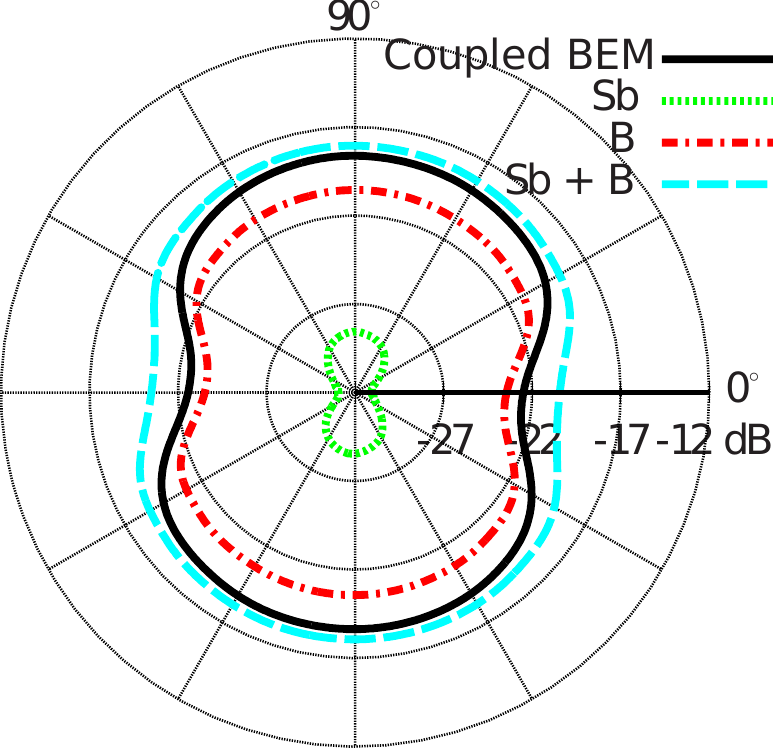}
	\caption{ Forward scattering TS as a function of angle $ \theta $ in the dorsal-{ventral} aspect 
	(left panel) and lateral aspect (right panel). 
	Swimbladder and fish body are considered as a gas-filled and weakly scattering objects, respectively. Both 
	contributions are taken into account in different ways. In the Coupled BEM approach (solid line) the interaction between 
	swimbladder and fish body is explicitely considered. The curve ``Sb'' shows exclusively the response of the swimbladder 
	(dotted line); the curve ``Sb+B'' shows the response of the swimbladder coherently added to the fish body (dashed line) 
	and the curve ``B'' shows exclusively the response of the fish-body (dotted-dashed line).
	}
	\label{fig_polar_fw} 
\end{figure}

The corresponding patterns for the forward scattering case are shown in Figure \ref{fig_polar_fw} where a curve ``B'',
representing exclusively the acoustic response from the fish body --calculated using the single scatterer' BEM approach--, has been added.

The left panel of Figure \ref{fig_polar_fw} displays the dorsal-ventral aspect pattern whereas the right panel shows the lateral aspect.
The forward scattering has a smooth behavior regards to the angle. 
In a weakly scattering situation, which occurs certainly for the fish body given the parameters used ($c_1=1.04 c_0$ 
and $\rho_1=1.04\rho_0$, c.f. Table \ref{Tabla_datos}), the forward response is almost independent of orientation because 
it is mainly a volume phenomenon.
For the dorsal{-ventral} and lateral aspects the model predicts practically the same patterns. The swimbladder response ``Sb'', which has two lobes under
both aspects, is remarkably smaller than all the others and it can be seen that the coherent sum ``Sb + B'' always exceeds the response obtained
from the model.
 
The minimum and maximum difference between coherent sum and the model are 0.84 and 2.08 dB (dorsal{-ventral} aspect) or 0.56 and 
2.11 dB (lateral aspect), respectively.

 ~~~~~~~~~~~~~~~~~~~~~~~~~~~~~~~~~~~~~~~~~~~~~~~~~~~~~~~~~~~~~~~~~~
\subsection{Near-field and internal field calculations}
\label{nf_calculations}

\begin{figure}[htb]
	\begin{minipage}[c]{0.32\textwidth}
		\includegraphics[scale=0.325]{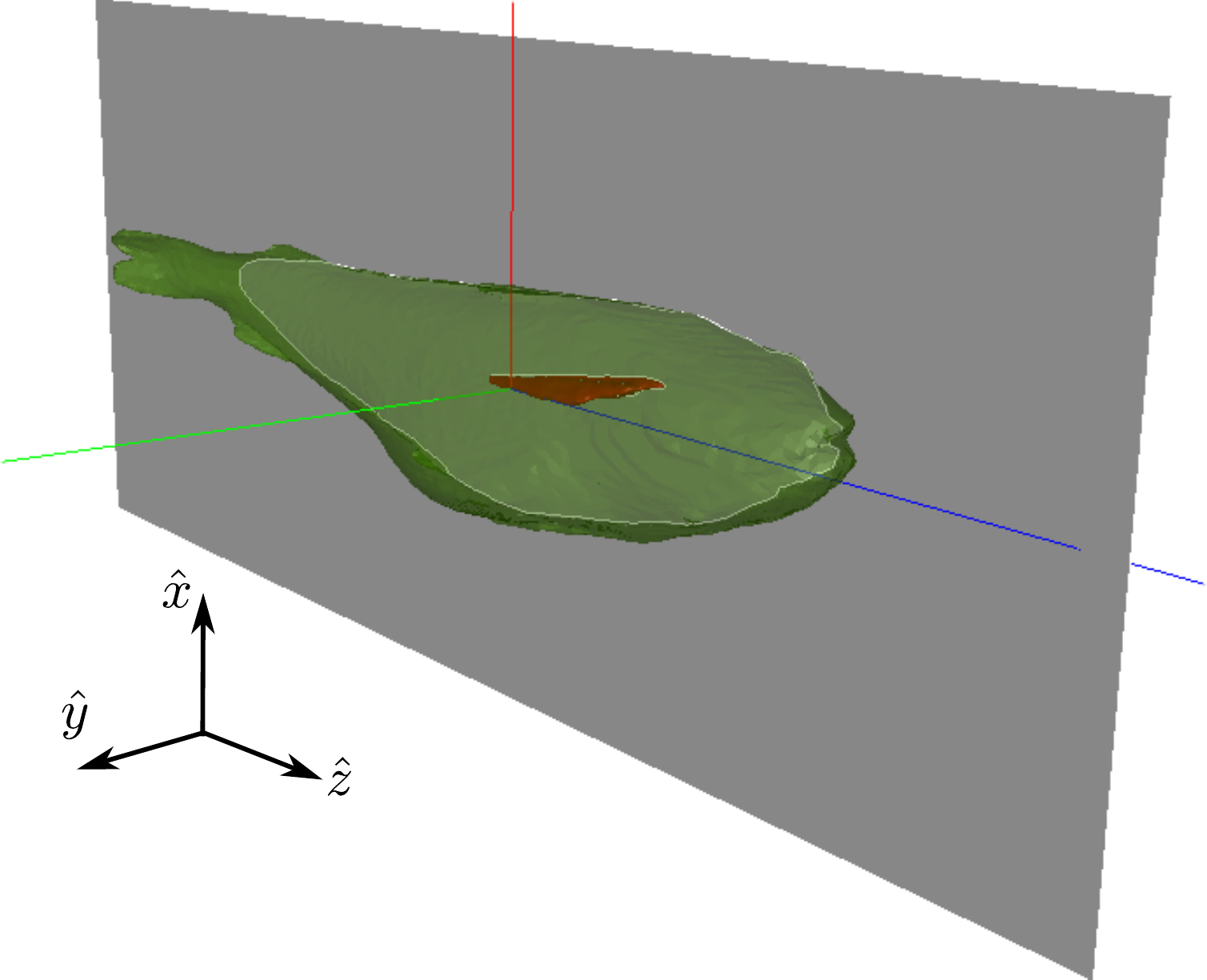}
	\end{minipage}
	\begin{minipage}[c]{0.05\textwidth}
		\hskip 3em
	\end{minipage}	
	\begin{minipage}[c]{0.60\textwidth}
		\includegraphics[scale=0.37]{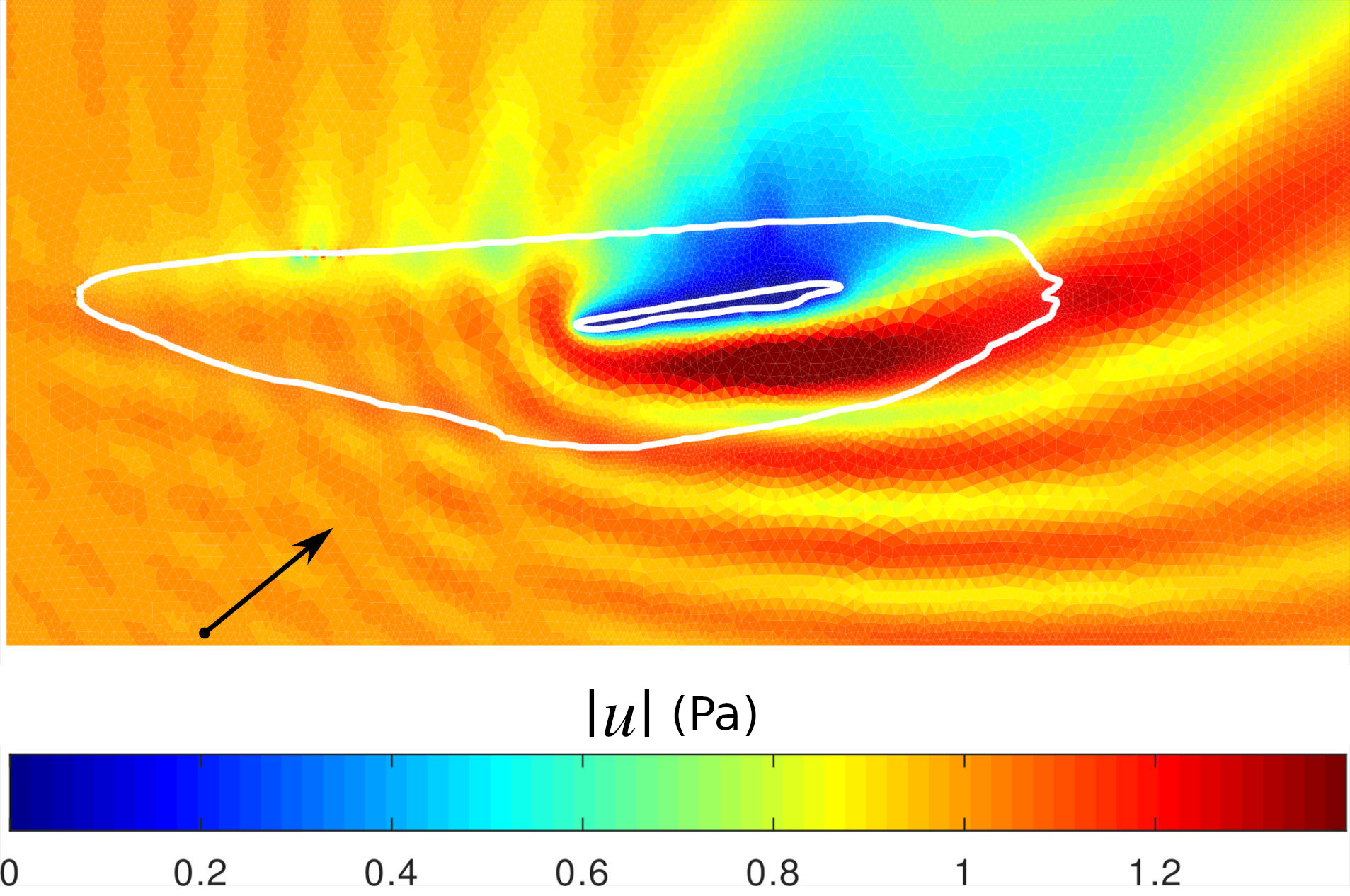}
	\end{minipage}
	\caption{  Diagram of fish and its intersection with a plane that cuts in two halfs the 
	parallelepid in which the swimbladder is inscribed (left). Evaluation of the absolute value of the near-field and the 
	internal fields over this plane (right). The resulting contour of the intersection between the plane and the 
	mesh-scatterers can be observed in both panels (solid line). The plane wave incidence direction is indicated by an 
	arrow.}
	\label{fig_scheme_nf}
\end{figure}

In Section \ref{verifications} the predictions derived from the formulated model were verified for the farfield case whereas in 
this Sub-Section, the near-field and internal fields are evaluated over a plane that cuts longitudinally in two halfs 
the parallelepid in which the swimbladder is inscribed. This plane is illustrated in the Figure \ref{fig_scheme_nf} 
(left), where the intersection curves between the fish body and the swimbladder with the plane can also be seen. 

The right panel of the Figure \ref{fig_scheme_nf} shows the the absolute value of the acoustic field on the plane. 
Again, the curves intersecting body and swimbladder with the plane, have been indicated. 
The plotted fields are $u = u_0 + \uinc$ in the exterior to the fish (the total field), 
$u_1$ in the fish flesh and $u_2$ inside the swimbladder.
An arrow indicating the incident field direction, a plane wave, has also been illustrated.

It can be clearly observed in the figure that the acoustic field verifies continuity across the different media. 
It should be noted that the changes in the field produced by the fish flesh are very weak while those due to the swimbladder lead to a high 
contrast (i.e. the field barely changes when entering the fish flesh from the exterior but this is not the case when the field enters 
the swimbladder from the fish internals) as it can be seen in the figure.
This state of affairs is undoubtedly \textit{tuned} by the impedance mismatch between the corresponding media; the data in Table \ref{Tabla_datos} prescribe a
high contrast between the swimbladder and the fish body and a minor one between the fish body and the surrounding water.

\section{Conclusions}
\label{conclusions}

A model has been formulated to describe the scattering produced by two penetrable objects immersed in an homogeneous medium. 
This model is based on a BIE formulation, subsequently discretized and solved numerically through a coupled BEM method. 
Verification of the good agreement between the model predictions and the results obtained by exact solutions in some benchmark 
cases was accomplished (Section \ref{verifications}). 

In particular, the model has been applied to compute the scattering produced by a fish of the hake type, \textit{Merluccius hubsi}
(Section \ref{application}) whose morphology was obtained from a CT tomography while the estimated values assumed for its physical properties
($c$ and $\rho$, sound speed and density, respectively) are {typical values reported in the literature for real fish.} 
From the standpoint of the model, fish flesh and swimbladder are considered as two penetrable and interacting objects.
However, it is emphasized that any extension of the model to enable the incorporation of other parts of the fish (e.g. spine) does 
not require further development and can be done following the same lines.

Results for backscattering and forward scattering TS have been obtained and compared with the individual responses of the swimbladder 
and the body as well as with its coherent sum (swimbladder and body contributions considered separately, i.e. without interaction 
between them). For the physical situation under consideration, it was found that the relative contributions of the swimbladder 
and the fish body vary depending of the scattering direction.

In the backscattering case it turned out that for the directions where the swimbladder presents its greater area to the incident wave, 
the coherent sum and the model do not show significant differences. On the other hand, in the case of forward scattering the behavior is totally 
different. The swimbladder is not dominant as in the previous case and the coherent sum always overestimates the interacting field
provided by the coupled BEM. In this scenario the body-swimbladder interaction should be taken into account explicitly since if it 
is not considered, errors up to 2 dB could appear in the calculated TS.

The model is also capable of computing the acoustic field within the fish as it is presented in Section \ref{nf_calculations} for a
particular plane. 

The correct modelling of the interaction between the various constituent parts of a fish, in the idealization that
each of them can be characterized by constant values of sound speed $c$ and density $\rho$, allows to progress in the further 
understanding of the acoustic response of individual specimens. The model can be useful for determining specific characteristics that 
allow to discriminate one species from another. Then, based on the knowledge of the most relevant characteristics 
that describe the acoustic scattering behavior, simplified models aimed to predict the response by schools of fish can be formulated.

The original computational codes are released and can be downloaded from one of the author's github repository, so that the reproducibility 
of the results obtained is guaranteed \cite{github_repo}.

\section*{Acknowledgments}
This work was supported by the Program of the Argentinian Ministry of Defense (PIDDEF 13/14), the Argentinian Navy and 
the National Council for Scientific and Technical Research (CONICET). The authors acknowledge the helpful contribution 
of Dr. Ariel Cabreira with the National Institute for Fisheries Research and Development (INIDEP) regarding the 
segmentation process for the computerized tomography scans.



\clearpage



\clearpage
\appendix

\section*{Appendix: BEM applied to a single fluid scatterer}
\label{singleObject}

The acoustic scattering by a single fluid object can be considered as a particular case of the two 
objects problem in which the inner object in Figure \ref{esquema} is suppressed. 
Therefore, the natural boundary conditions lead to 
\begin{equation*}
\begin{array}{rclr} 
	\uinc(x) + u_0(x) &=& u_1(x) & \text{for } x \in \Gamma_1 \\
	\\
	\displaystyle\frac 1 {\rho_0} \: \partial_n \uinc(x) + \frac 1 {\rho_0} \:\partial_n u_0(x) &=& 
	\displaystyle \frac 1 {\rho_1} \:\partial_n u_1(x) & \text{for } x \in \Gamma_1\\
	\label{1bodytrans_cond_single}
\end{array} 
\end{equation*}
The acoustic fields are expressed as 
\begin{equation}
\begin{array}{cl} 
	u_0(x) =& d_{01} \: K_0 [\:\psi_1\:](x) + s_{01} \: S_0[\:\phi_1\:](x) \\
	u_1(x) =& d_{11} \: K_1[\:\psi_1\:](x) + s_{11} \: S_1[\:\phi_1\:](x) 
	\label{1bodyfields_def}
\end{array}
\end{equation}
where the constants $d_{ij}$ and $s_{ij}$ are chosen as  
\begin{equation*}
	\begin{array}{c}
	d_{01} = \rho_0, \qquad s_{01} = \rho_0^2, \qquad	d_{11} = \rho_1,  \qquad s_{11} = \rho_1^2 .
	\end{array}
\end{equation*}

By straightforward calculations similar to the ones performed in Section \ref{modelformulation}, 
the discretized versions of densities $\psi_1$ and $\phi_1$ of  (\ref{1bodyfields_def}) are $\boldsymbol{ \psi^1 }$ 
and $\boldsymbol{ \phi^1 }$, respectively. They should satisfy the system 
\begin{equation} 
	\left(\boldsymbol{D}_1 + \alpha_{01} \boldsymbol{I} \right) \begin{pmatrix} \boldsymbol{ \psi^1 } \\
	\boldsymbol{ \phi^1 } \\
	\end{pmatrix} =
	\begin{pmatrix}
	\boldsymbol{ f } \\
	\boldsymbol{ g }\\
	\end{pmatrix} ,
	\label{1bodysistema}
\end{equation}
where $\alpha_{01}=(\rho_0+\rho_1)/2$, matrix $\boldsymbol{D}_1$, $\boldsymbol{ f }$ and $\boldsymbol{ g }$ depend only
on $\Gamma_1$. They are defined in Eqs. (\ref{matrizD1}) and (\ref{fgdef}).

The TS values are obtained by computing $f_\infty$, according to Eq. \eqref{finf_bs_discretized} 
where the $\boldsymbol{ \psi^1 }$ and $\boldsymbol{ \phi^1 }$ are now the solutions of the system in Eq. \eqref{1bodysistema}.

\end{document}